\documentclass[twocolumn, letter]{jpsj3}
\usepackage{txfonts}

\usepackage{graphicx}% Include figure files
\usepackage{dcolumn}% Align table columns on decimal point
\usepackage{bm}% bold math
\usepackage{color}

\title{Hall effect on nontrivial quadrupole order in quasi-kagome compound URhSn }

\author{Yusei Shimizu\thanks{E-mail address: yshimizu@issp.u-tokyo.ac.jp}$^{1,3}$, Arvind Maurya$^{2}$, 
 Yoshiya Homma$^{3}$, Motoi Kimata$^{4}$, Toni Helm$^{5}$, Ai Nakamura$^{3}$, \\
 Dexin Li$^{3}$, Atsushi Miyake$^{3}$, and Dai Aoki$^{3}$%\\
}
\inst{$^{1}$Institute for Solid State Physics (ISSP), University of Tokyo, Kashiwa, Chiba, 277-8581, Japan,  \\
$^{2}$Department of Physics, Mizoram University, Aizawl, Mizoram, 796004, India, \\
$^{3}$Institute for Materials Research (IMR), Tohoku University, Oarai, Ibaraki, 311-1313, Japan,  \\
$^{4}$Advanced Science Research Center, Japan Atomic Energy Agency, Tokai, Ibaraki, 319-1195, Japan.\\
$^{5}$Helmholtz-Zentrum Dresden-Rossendorf, Dresden 01328, Germany
}
\abst{
This study focuses on the transport properties of the quasi-kagome compound URhSn, which exhibits 
successive phase transitions at $T_{\rm C} = $16 K (ferromagnetic phase) and  $T_{\rm O} = $54 K (intermediate phase).
 A large anomalous Hall component is present along the easy-magnetization axis ($H ||$ $ [0001]$), and the Hall resistivity shows very complex temperature- and field-dependence, with  a sign reversal at low temperatures.
The Hall resistivity exhibits a nonlinear and unusual field-dependence.  Interestingly, there exists an unusual  Hall component   that is not proportional to the magnetic susceptibility  for $H$ $||$ $ [0001]$ in both the intermediate and ferromagnetic states. These results reveal unconventional transport properties of URhSn, providing important insights into nontrivial multipolar phases in $5f$-electron systems. 
}
\begin{document}
\maketitle

%%%%%%%%%%%%%%%%%%%%%%%%%%%%%%%%%%%%%%%%%%%%
%\section{Introduction}
%%%%%%%%%%%%%%%%%%%%%%%%%%%%%%%%%%%%%%%%%%%%

Uranium compounds exhibit unique physical properties, such as unconventional superconductivity \cite{Ott_PRL_1983, Stewart_PRL_1984, Palstra_PRL_1985, Schlabitz_ZPhysB_1986, Maple_PRL_1986, Ran_Science_2019, Aoki_JPSJ_2019}, non-Fermi-liquid behavior near the magnetic quantum phase-transitions \cite{Brando_RevModPhys_2016}, exotic Kondo effects \cite{Cox_AdvPhys_1998}, and multipole orders \cite{Suzuki_PRB_2000, Ishii_JPSJ_2012, Akazawa_JPSJ_1998, Tokiwa_JPSJ_2001, Walker_PRL_2006}, and have thus attracted much attention.  Among uranium compounds, for more than three decades, URu$_{2}$Si$_{2}$ has been intensively studied both experimentally and theoretically as an exotic material that exhibits a so-called \textit{hidden order} for which the order parameter has not been defined \cite{Mydosh_JPhysC_2020, Harima}.  
Recently, it was theoretically  proposed that chirality in the $5f$ electron state plays an important role in understanding the hidden order in uranium compounds \cite{Hoshino_PRL_2023, Hayami_JPSJ_2023}.

In URhSn, which has a distorted kagome structure (space group $\#189$: $P\bar{6}2m$) corresponding to a  hexagonal ZrNiAl-type structure, the uranium atoms do not have inversion symmetry and they consist a network of triangular lattices \cite{Dwight_JLessCommonMetals_1974}. This material undergoes successive phase transitions at $T_{\rm C} = $16 K and $T_{\rm O} = $54 K \cite{Paltra_JMMM_1987, Tran_JMMM_1991, Miranbet_JMMM_1995, Tran_JAlloyComp_1995,  Shimizu_PRB_2020}. The transition at lower temperature is a ferromagnetic transition, whereas the higher-temperature transition (intermediate phase) may not be related to  a magnetic order 
because the transition temperature $T_{\rm O}(B)$ is reinforced by applying magnetic fields, according to a recent study using single-crystalline URhSn \cite{Shimizu_PRB_2020}. The large entropy release of $R$ln3 at $T_{\rm O}$ suggests that this nontrivial order originates from the freezing of $5f$-electron degrees of freedom \cite{Shimizu_PRB_2020}. An early neutron diffraction study of URhSn suggested that no magnetic reflection is observed at temperatures between $T_{\rm C} = $ 16 K and $T_{\rm O} = $ 54 K \cite{Miranbet_JMMM_1995}. Additionally, previous M\"{o}ssbauer spectroscopy measurements suggested no internal field below $T_{\rm O} = $54 K \cite{Kruk_PRB_1997}.  Moreover, recent high-pressure experiments on URhSn revealed an exotic pressure-temperature phase diagram, in which a possible bi-critical point may exist at approximately 6.25 GPa \cite{Maurya_PRB_2021}.

In terms of the intermediate phase below $T_{\rm O}$, recent resonant x-ray scattering experiments on single-crystalline URhSn demonstrated that the transition at $T_{\rm O}$ is a $\bm{q}$ = 0 transition in which the translational symmetry of the crystal structure is maintained \cite{Tabata_JPSJ_2025}. Moreover, nuclear magnetic resonance (NMR) measurements revealed that the twofold symmetry of the internal magnetic field at the Sn site is preserved below $T_{\rm O}$ \cite{Tokunaga_PRL_}. These experimental findings  suggest  that the transition at $T_{\rm O}$ is an antiferroquadrupole (AFQ) transition with $P321$ symmetry \cite{Tabata_JPSJ_2025, Tokunaga_PRL_}. It is also theoretically argued that  chiral ($P321$) and polar ($P31m$)
  orders are possible  in the intermediate phase of URhSn   \cite{Kikuchi_JPSJ_2023, Ishitobi}. Ultrasound experiments have   revealed a clear softening of
 $C_{44}$  just above $T_{\rm O}$, which is consistent with the occurrence of the AFQ order    \cite{Tsuchida_JPS_2023}.

To understand the $5f$ electron state of the nontrivial order in URhSn,  we focus on the transport properties of URhSn using single-crystalline samples. Specifically, in URhSn, the occurrence of chiral quadrupole order below $T_{\rm O}$ has been proposed. Hence, identifying possible nontrivial transport properties that may occur owing to coupling between the conduction electrons and  $5f$ electron quadrupole order is worthwhile.

%%%%%%%%%%%%%%%%%%%%%%%%%%%%%%%%%%%%%%%%%%%%
%\section{Experimental Procedures}
%%%%%%%%%%%%%%%%%%%%%%%%%%%%%%%%%%%%%%%%%%%%

Herein, we report the results of  the Hall resistivity of URhSn using single-crystalline samples for easy-magnetization $[0001]$ axis. Single-crystalline URhSn samples were grown in a tetra-arc furnace under argon atmosphere using the Czochralski pulling method.  The samples were annealed at 800 $^\circ$C for five days. 
Laue x-ray photographs were checked  for the ingots and the sample was cut and polished for  Hall measurements. 
 We also fabricated a Hall bar (with a thickness of $\sim$30 $\mu$m) on URhSn using a focused-ion beam (FIB) technique.
  The Hall effect of  URhSn samples  were measured at low temperatures down to 1.7 K in magnetic fields up to 9 T using   a Physical Properties Measurement System (Dynacool, Quantum Design.).

%%%%%%%%%%%%%%%%%%%%%%%%%%%%%%%%%%%%%%%%%%%%
%\section{Results and Discussion}
%%%%%%%%%%%%%%%%%%%%%%%%%%%%%%%%%%%%%%%%%%%%

Figure 1(a) shows the temperature-dependence of the Hall resistivity of URhSn when a magnetic field is applied along the easy-magnetization axis ($H ||$ $[0001]$); the measurements were performed in the range of 70 K to 1.7 K. The electric current was applied along the $ [11\bar{2}0]$ direction, which is perpendicular to $[0001]$. A clear anomaly is observed below  $T_{\rm O}$ = 55 K. This anomaly shifts to higher temperatures with increasing magnetic field, consistent with the previously reported field-dependence of $T_{\rm O}$ \cite{Shimizu_PRB_2020}. A hump-like (maximum) anomaly also appears below $T_{\rm O}$: the Hall resistivity exhibits a positive maximum around 30 K at 2 T, around 40 K at 4 T, and around 50 K at 9 T. As the magnetic field increases,  the temperature of this Hall-resistivity maximum becomes increasingly higher. Furthermore, upon cooling, the sign of the Hall resistivity changes (e.g., the sign changes around 25 K under a field of 4 T).  A ferromagnetic transition occurs at 16 K; however, the transition becomes a crossover in magnetic fields, making it difficult to define a clear $T_{\rm C}$  for the magnetization easy axis.

When there is a large variation in the magnetoresistance and susceptibility with temperature or magnetic field,
accurate evaluation of the anomalous contribution of the Hall effect in bulk samples becomes difficult. If the Hall-voltage contacts ($V_{+}$ and  $V_{-}$) are slightly misaligned, the contribution of the magnetoresistance can affect the Hall-voltage signal. To precisely evaluate the effect of misalignment of the Hall voltage contacts, we also  measured the Hall resistivity using the FIB sample [Fig. 1(b)]. Figure 1(c) shows the temperature-dependence of the Hall effect of the FIB sample at 2 T.
%%%%%
 Here, we measured the Hall resistivity of the FIB sample with the current direction   $J$ $||$ $ [10\bar{1}0]$.
 We observed  behavior    similar to that of  the bulk sample measured with $J$ $||$ $ [11\bar{2}0]$, suggesting the absence of anisotropy within the hexagonal basal plane.
%%%%%%
 A clear hump-like anomaly below $T_{\rm O}$ and a sign change of $\rho_{\rm H} (T)$  are apparent,  and thus  these phenomena are essential for URhSn.

\color{black}
%%%%%%%%%%%%%%%%%%%%%%%%%%%%%%%%%%%%%%%%%%%%%%%%%%%%%%%%%%%%%%%%%%%%%%%%%%%%%%%%%%%%%%%%%%%%%%%%%%%

\begin{figure}[t]
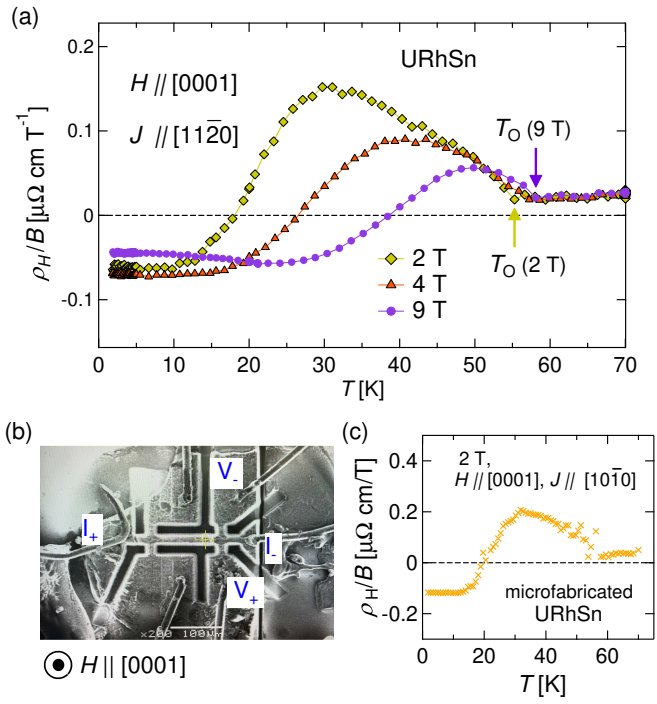

%%%%%%%%%%%%%%%%%%%%%%%
\begin{centering}
\includegraphics[width=8.9cm]{18158Fig1a.eps}
%%%%
\includegraphics[width=9.3cm]{18158Fig1bFig1c.eps}
\end{centering}
%\end{minipage} 
\caption{ (Color online)  (a) The temperature dependence of the Hall resisitivity of URhSn, measured at 
 2, 4, and 9  T applied along the  $H$ $||$ $[0001]$. The arrow indicates the nontrivial order at $T_{\rm O}$ in 9 T. (b)  The microfabricated URhSn sample with a Hall-bar geometry. 
 (c)  The temperature dependence of the Hall resisitivity of microfabricated URhSn, measured at  2 T applied along the  hexagonal  $[0001]$ axis. 
%%%%
 }
\end{figure}
%%%%%%%%%%%%%%%%%%%%%%%%%%%%%%%%%%%%%%%%%%%%%%%%%%%%%%%%%%%%%%%%%%%%%%%%%%%%%%%%%%%%%%%%%%%%%%%%%%

Below $T_{\rm O}$= 54 K, the Hall resistivity along the easy-magnetization axis exhibits a hump-like anomaly, 
  as indicated by arrows.
  This Hall-resistivity feature resembles that observed in the hidden-order phase of URu$_{2}$Si$_{2}$ at $T_{\rm HO}$ = 17.5 K \cite{Schoenes_PRB_1987, Bel_PRB_2004, Oh_PRL_2007}.
Previous studies on URu$_{2}$Si$_{2}$ attributed the anomalous hump-like behavior of the Hall resistivity below the hidden-order transition temperature $T_{\rm HO}$ to the opening of a gap on the Fermi surface \cite{Oh_PRL_2007}.  
%%%%%%%%%%%%%%%%%%%%%%%%%%%%%%%%%%%%%%%%%%%%%%%%%%
 In URhSn, as discussed later, the temperature dependence of the Hall effect around  the hump-like structure (with a maximum   around 39 K) cannot be explained solely by the anomalous Hall effect contribution.
 One possible scenario is a temperature-dependent change in the carrier concentration in URhSn due to the forming of a Fermi-surface gap below $T_{\rm O}$, but further investigations of  the ordinary Hall effect contribution is necessary to clarify this issue. 
%%%%%%%%%%%%%%%%%%%%%%%%%%%%%%%%%%%%%%%%%%%%%%%%%%
In URhSn, the field-dependence of  $T_{\rm O}$ indicates that the transition is reinforced by an applied magnetic field, which is characteristic of a quadrupolar-type order. As stated in the Introduction, recent resonant x-ray and NMR measurements revealed that this order corresponds to a $\bm{q} = $ 0 antiferroquadrupolar (AFQ) state with $P321$ symmetry \cite{Tabata_JPSJ_2025, Tokunaga_PRL_}. The present results provide an important reference for understanding the transport properties of uranium-based systems exhibiting multipole order with $\bm{q} =$ 0.

%%%%%%%%%%%%%%%%%%%%%%%%%%%%%%%%%%%%%%%%%%%%%%%%%%%%%%%%%%%%%%%%%%%%%%%%%%%%%%%%%%%%%%%%%%%%%%%%%%%

\begin{figure}[t]
%%%%%%%%%%%%%%%%%%%%%%%
\includegraphics[width=8.7cm]{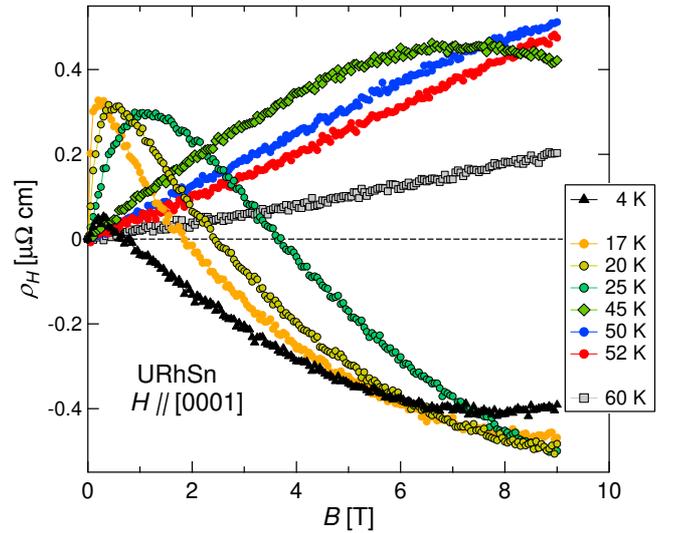}
%%%%%%%%%%%%
\caption{   (Color online)  
 $\rho_{\rm H}$ as a function of  magnetic field apllied  along $[0001]$  measured at $T$ = 4, 17, 20, 25, 45, 50, 52, and 60 K.
  }
 \label{f2}
\end{figure}
%%%%%%%%%%%%%%%%%%%%%%%%%%%%%%%%%%%%%%%%%%%%%%%%%%%%%%%%%%%%%%%%%%%%%%%%%%%%%%%%%%%%%%%%%%%%%%%%%%%

Figure 2 shows the magnetic-field-dependence of the Hall resistivity measured at various fixed temperatures with the application of a magnetic field along the easy-magnetization axis ($H $ $||$ $[0001]$). Here, URhSn is in the paramagnetic state at $T =$ 60 K and in the intermediate phase in the $T =$ 17-52 K range. In the paramagnetic state above $T_{\rm O}$, the Hall resistivity [$\rho_{\rm H}(B)$] exhibits an almost linear field-dependence. In contrast, $\rho_{\rm H}(B)$  exhibits a complex field-dependence in the intermediate phase between  $T =$ 17 and 52 K. At 45 K, the Hall resistivity increases linearly at low fields but shows a  
 broad maximum around 6-7 T.  
This maximum  shifts to lower magnetic fields upon cooling. At 25 K (20 K), the Hall resistivity reaches a positive maximum around 1 T (0.5 T). Thereafter, its sign changes near 4 T (3 T)  at 25 K (20 K). At 17 K, URhSn is still in the intermediate phase. However, this temperature is very close to the ferromagnetic Curie temperature ($T_{\rm C}$ = 16 K) and the Hall resistivity maximum occurs at lower field. At the lowest temperature, 4 K, in the ferromagnetic state,  the positive maximum disappears, and a small kink anomaly appears at weak fields: with increasing field, the sign of the Hall resistivity changes at 1 T, and $\rho_{\rm H}(B)$ is negative above 1 T. 
 A similar Hall resistivity curve has been observed in cuprate superconductors, where a Fermi-surface reconstruction accompanied by the opening of a gap occurs under  high fields \cite{Toni_2015}. When the opened gap is small, magnetic breakdown can occur at sufficiently high  fields, leading to extrema in the Hall resistivity as well as changes in its sign.
\color{black}

Interestingly, the maximum in the Hall resistivity $\rho_{\rm H}(B)$ curves observed for the intermediate phase is much larger than those for the ferromagnetic state. Note that in URhSn, the magnetization in the ferromagnetic state is much larger than that in the intermediate phase between $T_{\rm O}$ and $T_{\rm C}$
 \cite{Shimizu_PRB_2020}.
   This observation indicates that the anomalous Hall effect, which is  proportional to the magnetization, is expected to be much larger in the ferromagnetic state than in the intermediate phase. Nevertheless, the Hall effect in the intermediate phase is characterized by a larger Hall resistivity at low magnetic fields and a more pronounced field-dependence in URhSn.

Next, we discuss the anomalous Hall effect in URhSn. In general, the anomalous Hall effect is expressed as $\rho_{\rm H} = R_{\rm H} B = R_{0} B + R_{S} M$, where $R_{0}$ is the ordinary Hall coefficient, $R_{S}$ is the anomalous Hall coefficient, and $M$ is the magnetization \cite{HallEffect_ItsApplication}. Therefore, when $\rho_{\rm H}/B$ is plotted as a function of $M/B$, a linear relationship indicates that the Hall resistivity contains an anomalous Hall component.

Figures 3(a) and 3(b) show $R_{\rm H} = \rho_{\rm H}/B$ as a function of $M/B$, measured   (a) in the ferromagnetic state 
 ($T =$ 4 K) and (b)  in the intermediate phase ($T = $ 20 and 25 K). 
 At 4 K [Fig. 3(a)], $\rho_{\rm H}/B$ is proportional to $M/B$ from 0.4 to 4 T, indicating that the Hall resistivity is governed by the anomalous Hall effect in this region. 
 We find that the behavior of $R_{\rm H} =  \rho_{\rm H}/B \propto M/B$ is observed  when  $\rho_{\rm H}/B$
   becomes nearly zero and changes its sign. 
 In addition,   there exists an unusual property that   $R_{\rm H} = \rho_{\rm H}/B$ is not proportional to  $M/B$ above 4 T, showing  an  upturn behavior [Fig. 3(a)], although the magnetization is almost saturated in the ferromagnetic state  \cite{Shimizu_PRB_2020}.
   This behavior corresponds to the nonlinear behavior of Hall resistivity $\rho_{\rm H}(B)$ above 4 T (Fig. 2).
\color{black}

In the intermediate phase at $T=$ 25 K  [Fig. 3(b)], the proportional relationship between $\rho_{\rm H}/B$ and $M/B$ is limited to the field range of 3.3 $<B<$ 5.5 T (dashed line). 
 In the intermediate phase, 
$\rho_{\rm H}/B$ might be roughly explained by  $(M/B)^2$, possibly owing to the field-induced magnetic moments in the antiferroquadrupole order or spin-fluctuation effects.

%%%%%%%%%%%%%%%%%%%%%%%%%%%%%%%%%%%%%%%%%%%%%%%%%%%%%%%%%%%%%%%%%%%%%%%%%%%%%%%%%%%%%%%%%%%%%%%%%%%%%%%%%
\begin{centering}
\begin{figure}[t]
%%%%%%%%%%%%%%%%%%%%%%%
\includegraphics[width=9cm]{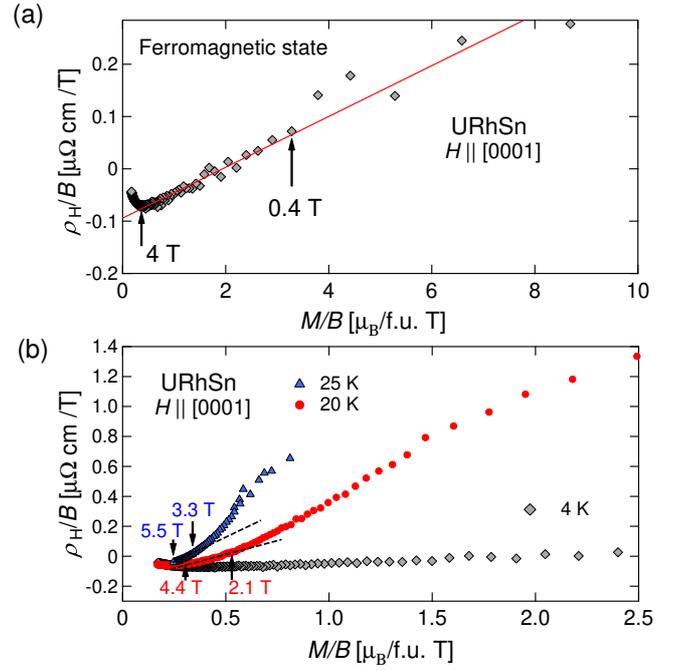}
%\end{minipage} 
%%%%%%%%%%%%
\caption{ (Color online) 
 $\rho_{\rm H}/B$ as a function of $M/B$ in magnetic fields along $[0001]$ 
 measured at    (a)  ferromagnetic state ($T$ = 4 K) and    (b) non-trivial ordered state   ($T$ = 20 and 25 K), respectively.
 The red line indicates the result of linear fitting between 0.4 and 4 T for data at 4 K, and the dashed  lines are guides
  where $\rho_{\rm H}/B $ is proportional to $ M/B$ in the  intermediate phase. 
 }
\end{figure}
\end{centering}
%%%%%%%%%%%%%%%%%%%%%%%%%%%%%%%%%%%%%%%%%%%%%%%%%%%%%%%%%%%%%%%%%%%%%%%%%%%%%%%%%%%%%%%%%%%%%%%%%%%%%%%%%

As mentioned  above, although the region in which the Hall resistivity ($\rho_{\rm H}/B$) is proportional to the magnetic susceptibility ($M/B$) is limited around 3-4 T, we shall attempt to decompose the anomalous Hall component in URhSn into contributions from skew scattering and the side-jump effect.
\color{black}
 In magnetic materials, the anomalous Hall coefficient ($R_{S}$)  can be expressed as the sum of the skew-scattering and side-jump effects \cite{Karplus_Luttinger_PR_1954}: $R_{S}(B) = a \rho + b \rho^2$, where $a$ and $b$ are  
  experimentally determined coefficients   and  $\rho$ is resistivity under magnetic fields. 
   The plot of $\rho_{\rm H}/B$ versus $\rho M$ [Fig.  4(a)] shows no proportional relationship, suggesting that skew scattering alone is insufficient for explaining $R_{S}$. Therefore, we take the side-jump  effect into account alongside the skew scattering. In this case, the Hall resistivity should satisfy the following relation: $R_{S}/\rho = (R_{\rm H} - R_{0} )B/\rho M = a + b \rho$. 
  %%%%
 Here, when evaluating the temperature dependence of the Hall resistivity measured at 4 T using 
  the above equation, the ordinary Hall coefficient ($R_{0}$) is assumed to be the value  obtained from the fitting in Fig. 3(a) ($R_{0} = - 0.09 $ $\mu \Omega$ cm/T)  at 4 K (ferromagnetic phase). 
  Note that this assumption would be  valid only when the anomalous Hall contribution is much larger than the ordinary  Hall contribution.
  Figure 4(b) shows a plot of $R_{S}/\rho$ versus $\rho $. A   proportional relationship holds over a wide temperature range (16-39 K), indicating that the Hall resistivity of URhSn at 4 T  may  be described by the skew-scattering and side-jump effects
   below the temperature at which the maximum anomaly occurs in $\rho_{\rm H}(T)$ [Fig. 1(a)]. 
   The change in the slope of  $(R -R_{0})B/ \rho M$ below $T_{\rm C }$ may suggest  a difference in scattering mechanisms between the ferromagnetic and intermediate phases.
     \color{black}
   A Hall resistivity with skew-scattering and side-jump contributions has also been observed in the itinerant metamagnet UCoAl, which has the same ZrNiAl-type crystal structure \cite{TDMatsuda_PRB_2000}.
   %%%%
  The temperature dependence of  $\rho_{\rm H}(T) $ above 39 K [Fig. 1(a)] in URhSn cannot be explained by the above analysis [Fig. 4(b)]. 
  In reality, the carrier density (ordinary Hall coefficient $R_{0}$)  may vary  with  temperature. 
  To explore the carrier properties in URhSn, a  more detailed  analysis using magnetization $M(B)$ curves and magnetoresistance will be required in the future.

We shall now discuss the origin of  Hall resistivity component that is not proportional to the magnetic susceptibility 
  in the intermediate phase below $\sim$ 2 T.
In low-field region, the maximum  anomaly in the Hall resistivity $\rho_{\rm H}(B)$ occurs simultaneously   in the intermediate phase ($T=$ 20, 25 K) as shown in Fig. 2. 
 This maximum shifts to lower fields upon cooling. At $T =$17 K, just above $T_{\rm C}$, the maximum of the Hall resistivity is still observed at lower fields. In the  ferromagnetic state, this maximum is significantly suppressed and becomes a kink at low field. 
 This evolution of the Hall resistivity maximum may be related to ferromagnetic fluctuation effects.

\color{black}

%%%%%%%%%%%%%%%%%%%%%%%%%%%%%%%%%%%%%%%%%%%%%%%%%%%%%%%%%%%%%%%%%%%%%%%%%%%%%%%%%%%%%%%%%%%%%%%%%%%%%%%%%
\begin{centering}
\begin{figure}[t]
%%%%%%%%%%%%%%%%%%%%%%%
\includegraphics[width=9cm]{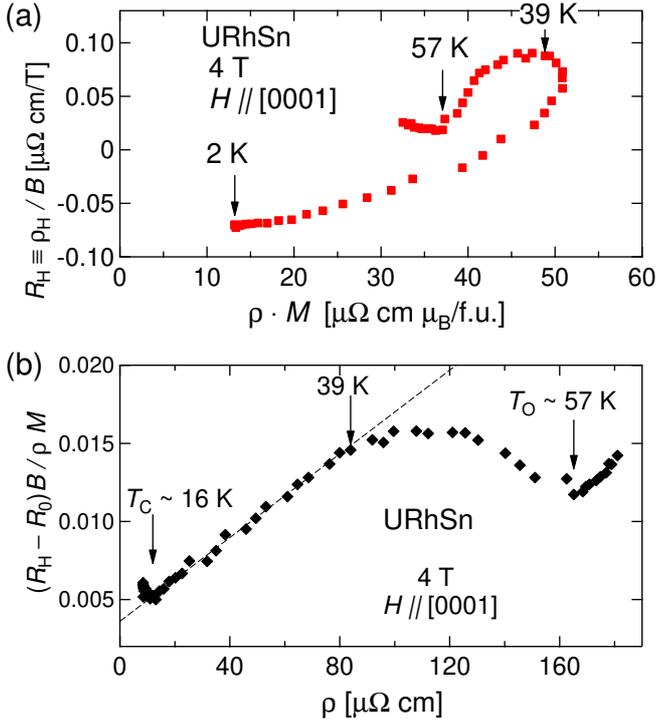}
%%%%%%%%%%%%%%%%%%%%%%%%%%%%%
\caption{  (Color online)   (a) $R_{\rm H} \equiv \rho_{\rm H}/ B$ vs. $ \rho M$ plot for $\rho_{\rm H}(T)$ data measured at 4 T. 
(b) $(R_{\rm H} - R_0 ) B/ \rho M$ of URhSn  as a function of $\rho $ applied  along the easy-magnetization axis  $[0001]$  measured  at 4 T,  where $R_0 = -0.09$ [$\mu \Omega $cm/T] was assumed. Here, the dashed line indicates the guide to the eyes. 
 }
 \label{f2}
\end{figure}
\end{centering}
%%%%%%%%%%%%%%%%%%%%%%%%%%%%%%%%%%%%%%%%%%%%%%%%%%%%%%%%%%%%%%%%%%%%%%%%%%%%%%%%%%%%%%%%%%%%%%%%%%%

We propose  the another origin of the non-proportionality between the Hall resistivity and magnetic susceptibility 
  below 2 T in URhSn. 
  The intermediate phase of URhSn is most likely explained by a $\bm{q}  =$ 0 antiferroquadrupole  order with chiral $P321$ and  polar $P31m$     space groups \cite{Tabata_JPSJ_2025, Tokunaga_PRL_}. 
The order parameters  of $P321$ and $P31m$ symmetries  are  antiferroquadrupole  $O_{yz}$ and $O_{zx}$  \cite{Tabata_JPSJ_2025, Tokunaga_PRL_}. 
 Here, the $z$-axis is defined as the $[0001]$ axis, the hexagonal $[10\bar{1}0]$ direction as the $y$-axis, and the direction perpendicular to the $y$-axis  as the $x$-axis. 
 When a magnetic field is applied along $H$ $||$ $[0001]$, 
  magnetic dipole moments $J_{y}$ and  $J_{x}$ are induced 
   at the uranium sites within the hexagonal basal  plane   for  the antiferroquadrupole  $O_{yz}$ ($P321$) and $O_{zx}$ ($P31m$)  \cite{Tokunaga_PRL_, Shiina_JPSJ_1997}. 
     For  the antiferroquadrupole  $O_{yz}$ ($P321$)  an all-in-/all-out-like spin structure is expected at uranium sites, whereas  
   a vortex-like spin structure is anticipated   for  $O_{zx}$ ($P31m$) \cite{Tabata_JPSJ_2025, Ishitobi}. 
 For  $H$ $||$  $[0001]$ $||$  $z$, a paramagnetic magnetization component proportional to  magnetic dipole  $J_{z}$ is also induced at the uranium sites. 
 Consequently, microscopic magnetic moments that are tilted relative to both the $[0001]$ axis and the hexagonal basal  plane can be induced at each uranium site \cite{Ishitobi}. 
 %%%%
 Because these induced magnetic moments become non-coplanar, scalar spin chirality can arise: $\chi_{123} \propto \bm{m}_{1} \cdot \bm{m}_{2} \times \bm{m}_{3}$, where $\bm{m}_{i}$ ($i = $1, 2, 3) are the induced  momentums at the triangular uranium sites.

  When conduction electrons interact with localized $5f$ moments possessing such scalar spin chirality, a Berry phase can emerge, leading to a topological Hall effect in URhSn. Topological Hall effects  have also been observed in the pyrochlore compound  Nd$_{2}$Mo$_{2}$O$_{7}$ \cite{Taguchi_Science_2001}, in skyrmion compounds such as  MnSi \cite{Neubauer_PRL_2009, Lee_PRL2009}, and MnGe \cite{Kanazawa_PRL_2011}, and in the non-coplanar antiferromagnetic order in UCu$_{5}$ \cite{Ueland_NatureComm_2012}.
  %%%%%%%%%%%%%%%%%%%%%%%%
  In particular, a similar peak anomaly  in Hall resistivity [$\rho_{\rm H}(B)$] has been reported  in  MnSi \cite{Neubauer_PRL_2009, Lee_PRL2009}  and  UCu$_{5}$  \cite{Ueland_NatureComm_2012}. 
  In URhSn, 
   with increasing field, the  $J_{y}$ ($J_{x}$)  induced for the $O_{yz}$ ($O_{zx}$) quadrupoles within the hexagonal plane may become saturated at a certain field.  
   In high magnetic fields, $J_{z}$ should be much larger than  $J_{x}$, $J_{y} $, i.e., $J_{x}$, $J_{y} $ $\ll$ $ J_{z}$. Hence, the topological Hall component may reach a maximum at a certain field.
     This may be the origin of the maximum in the Hall resistivity observed around 1-2 T.
    The scalar spin chirality may emerge in  both the $P321$ and $P31m$ cases, and thus   it remains unclear which  order parameter is more plausible for  explaining   the Hall resistivity results in URhSn. 
  Also,  at this stage,  we cannot rule out a possibility  that (thermal) spin  fluctuations may induce such maximum in 
  $\rho_{\rm H}(B)$.
 To verify the scalar spin chirality in URhSn, further theoretical studies and  neutron scattering experiments under magnetic fields are prospectively necessary.

Finally, we briefly discuss the importance of understanding the electronic structure of URhSn. 
Because the crystal structure of URhSn lacks inversion symmetry (space group $\#189$), Fermi-surface splitting owing to spin-orbit coupling is expected under magnetic fields \cite{Onuki_NonCentrosymmetric, Zhang_PRB_2014}. 
The Fermi-surface splitting and the asymmetry of the opposite spin textures for up-spin and down-spin electrons 
under high magnetic fields may induce a Berry phase in momentum space. 
 %%%%%
 As seen in Fig. 2, in the  ferromagnetic ($T$ = 4 K) and intermediate  ($T$ = 17, 20, 25 K) states, $\rho_{\rm H}(B)$   showing an unusual nonlinear curves above 4 T. 
 This behavior might be explained by the unusual band structure  in this noncentrosymmetric compound.
To deepen the electronic structure in URhSn,  more precise Hall measurements  with microfabricated devices
 and quantum oscillation experiments  are  necessary. 
\color{black}

%%%%%%%%%%%%%%%%%%%%%%%%%%%%%%%%%%%%%%%%%%%%
%\section{Summary}
%%%%%%%%%%%%%%%%%%%%%%%%%%%%%%%%%%%%%%%%%%%%

In summary, we examined the Hall effects in the quasi-kagome compound URhSn with the nontrivial $\bm{q} =$ 0 order,
which possibly arises from the chiral antiferroquadrupole order. 
 For $H$ $||$  $[0001]$,  a hump anomaly is observed below $T_{\rm O}$ (intermediate phase), 
 accompanied by a clear  sign change at low temperatures. 
 \color{black}
This unusual transport phenomenon was also observed  using a Hall-bar sample prepared via  FIB  microfabrication. Moreover, for $H$ $||$$[0001]$,   we observed a peculiar Hall contribution that is not proportional to the   magnetic   susceptibility  both in the ferromagnetic and intermediate phases in URhSn.  
 To explain the Hall effect in URhSn, the skew scattering and side-jump effect alone are not sufficient,
  and there might be topological Hall effect in this non-centrosymmetric material. 
\color{black}
These results  provide important insight for deeper understanding of hidden orders in enigmatic $5f$-electron systems with chiral multipole orders.

\color{black}

%\begin{acknowledgments}
We are grateful to H. Harima, K. Hattori, T. Ishitobi, Y. Tokunaga, and C. Tabata  for their valuable discussions. The present study was supported by Grants-in-Aid KAKENHI (No. JP20K03851, JP22KK0224, JP23K03314, and JP23H04870) from the Ministry of Education, Culture, Sports, Science and Technology (MEXT) of Japan.

%%%%%%%%%%%%%%%%%%%%%%%%%%%%%%%%%%%%%%%%%%%%%%%%%%%%%%%%%%%%%%%%%%%%%%%%%%%%%%%%%%%%%%%%%%%%%%%%%%%%%%%%%%%%%%%%%%%%%%%%%%%%
%\bibliography{apssamp}% Produces the bibliography via BibTeX. 

\begin{thebibliography}{99}
%%%%%%%%%%%%%%%%%%%%%%%%%


%%%%% \r{a} \^a \.{e}  {\AA}
%%%%%%%%%%%%%%%%%%%%%%%%%%%%%%%%%%%%%%%%%%%%%%%%%%%%%%%%%%%%%%%%%%%%%%%%%%%%%%%%%%%%%%%%%%%%%%%%%%%%%%%%%%%%%
%%%%%%%%%%%%%%%%%%%%%%%%%%%%%%%%%%%%%%%%%%%%%%%%%%%%%%%%%%%%%%%%%%%%%%%%%%%%%%%%%%%%%%%%
\bibitem{Ott_PRL_1983} H. R. Ott, H. Rudigier, Z. Fisk, and J. L. Smith, Phys. Rev. Lett. {\bf50}, 1595 (1983).
%%%%%%%%%%%%%%%%%%%%%%%%%%%%%%%%%%%%%%%%%%%%%%%%%%%%%%%%%%%%%%%%%%%%%%%%%%%%%%%%%%%%%%%%
\bibitem{Stewart_PRL_1984} G. R. Stewart, Z. Fisk, J. O. Willis, and J. L. Smith, Phys. Rev. Lett. {\bf52}, 679 (1984).
%%%%%%%%%%%%%%%%%%%%%%%%%%%%%%%%%%%%%%%%%%%%%%%%%%%%%%%%%%%%%%%%%%%%%%%%%%%%%%%%%%%%%%%%
\bibitem{Palstra_PRL_1985}   T. T. M. Palstra, A. A. Menovsky, J.van den Berg, A. J. Dirkmaat,
P. H.Kes, G. J. Nieuwenhuys, and J. A. Mydosh, Phys. Rev. Lett. {\bf55}, 2727 (1985). 
%%%%%%%%%%%%%%%%%%%%%%%%%%%%%%%%%%%%%%%%%%%%%%%%%%%%%%%%%%%%%%%%%%%%%%%%%%%%%%%%%%%%%%%%%%%%%
\bibitem{Schlabitz_ZPhysB_1986} W. Schlabitz, J. Baumann, B. Pollit, U. Rauchschwalbe, H. M. Mayer, U. Ahlheim, and C. D. Bredl, Z. Phys. B  {\bf62}, 171 (1986).
%%%%%%%%%%%%%%%%%%%%%%%%%%%%%%%%%%%%%%%%%%%%%%%%%%%%%%%%%%%%%%%%%%%%%%%%%%%%%%%%%%%%%%%%%%%%
\bibitem{Maple_PRL_1986} M. B. Maple, J. W. Chen, Y. Dalichaouch, T. Kohara, C. Rossel, M. S. Torikachvili, M.W. McElfresh, and J. D. Thompson, Phys. Rev. Lett. {\bf56}, 185 (1986).


\bibitem{Ran_Science_2019} S. Ran, C. Eckberg, Q.-P. Ding, Y. Furukawa, T. Metz, S. R. Saha, I.-L. Liu, M. Zic, H. Kim, J. Paglione, and N. P. Butch, Science 365, 684 (2019).

\bibitem{Aoki_JPSJ_2019} D. Aoki, A. Nakamura, F. Honda, D. X. Li, Y. Homma, Y. Shimizu, Y. J. Sato, G. Knebel, J.-P. Brison, A. Pourret, D. Braithwaite, G. Lapertot, Q. Niu, M. Valiska, H. Harima, and J. Flouquet, J. Phys. Soc. Jpn. 88, 043702 (2019).

%%%%%%%%%%%%%%%%%%%%%%%%%%%%%%%%%%%%%%%%%%%%%%%%%%%%%%%%%%%%%%%%%%%%%%%%%%%%%%%%%%%%%%%%
\bibitem{Brando_RevModPhys_2016} M. Brando, D. Belitz, F. M. Grosche, and T. R. Kirkpatrick, Rev. Mod. Phys. {\bf88}, 025006 (2016).
 and references cited therein. 
%%%%%%%%%%%%%%%%%%%%%%%%%%%%%%%%%%%%%%%%%%%%%%%%%%%%%%%%%%%%%%%%%%%%%%%%%%%%%%%%%%%%%%%%
\bibitem{Cox_AdvPhys_1998} D. L. Cox and A. Zawadowski, Adv. Phys. {\bf47}, 599 (1998). 
 and references cited therein. 
%%%%%%%%%%%%%%%%%%%%%%%%%%%%%%%%%%%%%%%%%%%%%%%%%%%%%%%%%%%%%%%%%%%%%%%%%%%%%%%%%%%%%%%%


%%%%UCu2Sn%%%%%%%%%%%%%%%%%%%%%%%%%%%%%%%%%%%%%%%%%%%%%%%%%%%%%%%%%%%%%%%%%
\bibitem{Suzuki_PRB_2000}  T. Suzuki, I. Ishii, N. Okuda, K. Katoh, T. Takabatake, T. Fujita,  A. Tamaki,  Phys. Rev. B  {\bf62},  49  (2000). 
%%%%%%%
%%%%%%%%%%%%%%%%%%%%%%%%%%%%%%%%%%%%%%%%%%%%%%%%%%%%%%%%%%%%%%%%%%%%%%%%%%%%%%%%
\bibitem{Ishii_JPSJ_2012} I. Ishii, K. Katoh, T. Takabatake, S. Hashio, A. Tamaki, and T. Suzuki, J. Phys. Soc. Jpn. {\bf81}, 024602 (2012). 


%%%%%%%% UNiSn %%%%%%%%%%%%%%%%%%%%%%%%%%%%%%%%%%%%%%%%%%%%%%%%%%%%%%%%%%%%%%%%%%%%%%%%%%%%%%%%%%%%%
\bibitem{Akazawa_JPSJ_1998} T. Akazawa, T. Suzuki, H. Goshima, T. Tahara, T. Fujita, T. Takabatake, and H. Fujii, J. Phys. Soc. Jpn. {\bf67},
 3256 (1998).
%%%%%%%%%%%%%%%%%%%%%%%%%%%%%%%%%%%%%%%%%%%%%%%%%%%%%%%%%%%%%%%%%%%%%%%%%%%%%%%%
\bibitem{Tokiwa_JPSJ_2001} Y. Tokiwa, K. Sugiyama, T. Takeuchi, M. Nakashima, R. Settai, Y. Inada, Y. Haga, E. Yamamoto, K. Kindo, H. Harima, and Y. Onuki, J. Phys. Soc. Jpn. {\bf70}, 1731 (2001).
%%%%%%%%%%%%%%%%%%%%%%%%%%%%%%%%%%%%%%%%%%%%%%%%%%%%%%%%%%%%%%%%%%%%%%%%%%%%%%%%%%%%%
\bibitem{Walker_PRL_2006} H. C. Walker, K. A. McEwen, D. F. McMorrow,  S. B. Wilkins, F. Wastin, E. Colineau,  and D. Fort, Phys. Rev. Lett. {\bf97}, 
 137203 (2006). 
%%%%%%%%%%%%%%%%%%%%%%%%%%%%%%%%%%%%%%%%%%%%%%%%%%%%%%%



\bibitem{Mydosh_JPhysC_2020}  J. A. Mydosh, P. M. Oppeneer, and P. S. Riseborough, J. Phys. Condens. Matter {\bf32}, 143002 (2020). 

\bibitem{Harima} H. Harima, SciPost Phys. Proc. {\bf11}, 006 (2023).


\bibitem{Hoshino_PRL_2023} S. Hoshino, M.-T. Suzuki, and H. Ikeda, Phys. Rev. Lett. 130, 256801 (2023).


\bibitem{Hayami_JPSJ_2023}  S. Hayami and H. Kusunose, J.  Phys. Soc. Jpn. 92, 113704 (2023).




%%%%%%%%%%%%%%%%%%%%%%%%%%%%%% URhSn %%%%%%%%%%%%%%%%%%%%%%%%%%%%%%%%%%%%%
%%%%%%%%%%%%%%%%%%%%%%%%%%%%%%%%%%%%%%%%%%%%%%%%%%%%%%%%%%%%%%%%%%%%%%%%%%%%%%%%
%%%%%% URhSn, %%%%%
\bibitem{Dwight_JLessCommonMetals_1974} A. E. Dwight, J. Less.-Common Metals, {\bf34}, 279 (1974).


\bibitem{Paltra_JMMM_1987} T. T. M. Paltra, G.J. Nieuwenhuys,  R.F.M. Vlastuin, J. van den Berg,  J. A. Mydosh, J. Magn. Magn. Mat. {\bf67}, 331 (1987).
%%%%%%%%%%%%%%%%%%%%%%%%%%%%%%%%%%%%%%%%%%%%%%%%%%%%%%%%%%%%%%%%%%%%%%%%%%%%%%%%
\bibitem{Tran_JMMM_1991} V. H. Tran and R. Tro\'{c}, J. Mag. Mag. Mat. {\bf102}, 74 (1991).
%%%%% URhSn, C/T and neutron %%%%%%%%%%%%%%%%%%%%%%%%%%%%%%%%%%%%%%%%%%%%%%%%%%%%%%%%%%
\bibitem{Miranbet_JMMM_1995} F. Mirambet,  B. Chevalier, L. Fourn\`es, J. Ferreira da Silva, M. A. Frey Ramos,   T. Roisnel, J. Mag. Mag. Mat. {\bf140-144}, 1387 (1995). 

%%%%%% URhSn, UCoSn, etc AF &FM %%%%%%%%%%%%%%%%%%%%%%%%%%%%%%%%%%%%%%%%%%%%%%%%%%%%%%
\bibitem{Tran_JAlloyComp_1995} V. H. Tran, R. Tro\'{c}, and D. Badurski, J. Alloys and Comps {\bf219}, 285 (1995).
%%%%%%%%%%%%%%%%%%%%%%%%%%%%%%%%%%%%%%%%%%%%%%%%%%%%%%%%%%%%%%%%%%%%%%%%%%%%%%%%%%%%%%%


\bibitem{Shimizu_PRB_2020}  Y. Shimizu,  A.  Miyake, A. Maurya,  F. Honda, A. Nakamura,  Y. J. Sato,  D. X. Li, 
Y.  Homma,  M.  Yokoyama, Y. Tokunaga,  M.  Tokunaga,  and D. Aoki, Phys. Rev. B {\bf102}, 134411 (2020). 

%%%%% Mossbauer for URhSn
\bibitem{Kruk_PRB_1997}  R. Kruk,  R. Kmie\'{c}, K.   {\L}atka, K. Tomala, R. Tro\'{c}, and V. H. Tran, Phys. Rev. B {\bf55}, 5851  (1997). 
%%%%%


\bibitem{Maurya_PRB_2021} A. Maurya,  D. Bhoi,  F.  Honda,  Y. Shimizu, A. Nakamura, Y. J. Sato,  D. X. Li,  Y.  Homma, M. . Sathiskumar,  J.  Gouchi, Y.  Uwatoko, and D. Aoki, Phys. Rev. B {\bf104}, 195119 (2021).  
 

\bibitem{Tabata_JPSJ_2025} C. Tabata, F. Kon, R. Hibino, Y. Shimizu, H. Amitsuka, K. Kaneko, Y. Homma, D. Aoki, and H. Nakao, J. Phys. Soc. Jpn. {\bf94}, 083701 (2025).

\bibitem{Tokunaga_PRL_} Y.Tokunaga, T. Ishitobi, H. Sakai, S.Kambe,  H. Harima,  A. Nakamura, A. Maurya, 
D. Li, Y. Homma, F. Honda, D. Aoki, and Y. Shimizu, JPS Spring Meeting,
 25aH1-12 (2023), submitted to Phys. Rev. Lett. 


\bibitem{Kikuchi_JPSJ_2023} H. Kusunose and J. Kikuchi, J. Phys. Soc. Jpn.  {\bf93}, 074701 (2024).

\bibitem{Ishitobi} T. Ishitobi and K. Hattori, arXiv:2502.13977 (2025).


\bibitem{Tsuchida_JPS_2023} K. Tsuchida, R. Hibino, M. Matsuda, H. Hidaka, T. Yanagisawa, H. Amitsuka, C. Tabata, and Y. Shimizu,  JPS Annual Meeting , 18aA205 (2023).


%%%%%%%%%%%%%%%%%%%%%%%%%%%%%%%%%%%%%%%%%%%%%%%%%%%%%%%%%%%%%%%%%%%%%%%%%%%%%%%%%%%%%%%%


\bibitem{Schoenes_PRB_1987} J. Schoenes, C. Scho..nenberger, J. J. M. Franse, and A. A. Menovsky, Phys. Rev. B {\bf35}, 5375 (1987). 

\bibitem{Bel_PRB_2004} R. Bel,  H. Jin, K. Behnia,  J. Flouquet,  and P. Lejay, Phys. Rev. B  {\bf70}, 220501(R) (2004).

\bibitem{Oh_PRL_2007} Y. S. Oh,  K. H. Kim,  P. A. Sharma,   N. Harrison,  H. Amitsuka, and J. A. Mydosh, Phys. Rev. Lett. {\bf 98}, 016401 (2007).


\bibitem{Toni_2015} T. Helm, M. V. Kartsovnik, C. Proust,  B. Vignolle,  C. Putzke,  E. Kampert,  I. Sheikin,  E.-S. Choi,  J. S. Brooks,  N. Bittner,  W. Biberacher,  A. Erb,  J. Wosnitza,  and R. Gross, Phys. Rev. B 92, 094501 (2015). 

\bibitem{HallEffect_ItsApplication} C. L. Chien and C. R. Westgate, The Hall Effect and Its Application, Springer Science + Business Media, New York (1980). 




\bibitem{Karplus_Luttinger_PR_1954} R. Karplus and J. M.  Luttinger, Phys. Rev. {\bf95}, 1154 (1954). 


\bibitem{TDMatsuda_PRB_2000} T. D. Matsuda, H. Sugawara, Y. Aoki, H. Sato, A. V. Andreev, Y. Shiokawa, V. Sechovsky and L. Havela,  Phys. Rev. B {\bf62}, 13852 (2000).



\bibitem{Shiina_JPSJ_1997} R. Shiina, H. Shiba, and P. Thalmeier, J. Phys. Soc.  Jpn.  {\bf66}, 1741 (1997). 



\bibitem{Taguchi_Science_2001} Y. Taguchi,  Y. Oohara,  H. Yoshizawa,  N. Nagaosa, Y. Tokura, Science {\bf291}, 2573 (2001).

\bibitem{Neubauer_PRL_2009} A. Neubauer,  C. Pfleiderer,  B. Binz,  A. Rosch,  R. Ritz,  P. G. Niklowitz,  and P. B\"{o}ni, Phys. Rev. Lett.  {\bf102}, 186602 (2009).

\bibitem{Lee_PRL2009} M. Lee, W. Kang, Y. Onose, Y. Tokura, and N. P. Ong, Phys. Rev. Lett.  {\bf102}, 186601 (2009).


\bibitem{Kanazawa_PRL_2011} N. Kanazawa, Y. Onose,  T. Arima, D. Okuyama, K. Ohoyama, S. Wakimoto, K. Kakurai, 
S. Ishiwata,  and Y. Tokura, Phys. Rev. Lett.  {\bf106}, 156603 (2011).

%%%%%% UCu5 %%%%%%%%%%%%%%%%%%%%%%%%%%%%%%%%%%%%%%%%%%%%%%%%%%%%%%%%%%%%%%%%%%%%%%%%%%%%%%%%
\bibitem{Ueland_NatureComm_2012} B. G. Ueland, C. F. Miclea, Y. Kato, O, Ayala-Valenzuela, R. D. MacDonald, R. Okazaki, P. H. Tobash, M. A. Torrez, F. Ronning, R. Movshovich,  Z. Fisk, E. D. Bauer, I. Martin, and J. D. Thompson, Nat. Commun., {\bf10}, 1038 (2012). 



\bibitem{Onuki_NonCentrosymmetric} Y. $\overline{\rm{ O} }$nuki and R. Settai, in Non-Centrosymmetric Superconductors, ed. 
E. Bauer and M. Sigrist (Springer, Heidelberg, 2012) Lecture Notes in Physics, Vol. 847, Ch. 3.

\bibitem{Zhang_PRB_2014}  X. Zhang, Q. Liu, J-W. Luo, A. J. Freeman,   and A. Zunger, Nat. Phys. {\bf 10} (2014). 

%\begin{thebibliography}{9}
%\bibitem{jpsj} The abbreviation for JPSJ must be ``J. Phys. Soc. Jpn." in the reference list.
%\bibitem{instructions} More abbreviations of journal titles are listed in ``Instructions for Preparation of Manuscript".
\end{thebibliography}
%%%%%%%%%%%%%%%%%%%%%%%%%%%%%%%

\end{document}